\documentstyle[12pt]{article}
\input psfig.sty

\begin{document}
\baselineskip=17pt
\pagestyle{plain}
\setcounter{page}{1}

\begin{titlepage}

\begin{flushright}
CLNS-98/1554 \\
hep-th/9804007
\end{flushright}
\vspace{10 mm}

\begin{center}
{\huge Deriving N=2 S-dualities from Scaling}\\
\vspace{5mm}
{\huge for Product Gauge Groups}
\vspace{5mm}
\end{center}
\vspace{10mm}
\begin{center}
{\large Philip C. Argyres and Alex Buchel}\\
\vspace{3mm}
{\it Newman Lab., Cornell University, Ithaca NY 14853}\\
argyres,buchel@mail.lns.cornell.edu
\end{center}
\vspace{15mm}
\begin{center}
{\large Abstract}
\end{center}
\noindent
S-dualities in scale invariant $N=2$ supersymmetric field theories
with product gauge groups are derived by embedding those theories in
asymptotically free theories with higher rank gauge groups.  S-duality
transformations on the couplings of the scale invariant theory follow
from the geometry of the embedding of the scale invariant theory in
the Coulomb branch of the asymptotically free theory.
\vspace{1cm}
\begin{flushleft}
April 1998
\end{flushleft}
\end{titlepage}
\newpage
\renewcommand{\baselinestretch}{1.1}  


\newcommand{\ATMP}{Adv.\ Theor.\ Math.\ Phys.\ }
\newcommand{\JHEP}{J.H.E.P.\ }
\newcommand{\NP}{Nucl.\ Phys.\ }
\newcommand{\PL}{Phys.\ Lett.\ }
\newcommand{\PR}{Phys.\ Rev.\ }
\newcommand{\PRL}{Phys.\ Rev.\ Lett.\ }
\newcommand{\CMP}{Commun.\ Math.\ Phys.\ }

\newcommand{\be}{\begin{equation}}
\newcommand{\ee}{\end{equation}}
\newcommand{\bea}{\begin{eqnarray}}
\newcommand{\eea}{\end{eqnarray}}

\newcommand{\cc}{{\cal C}}
\newcommand{\cg}{{\cal G}}
\newcommand{\mm}{{\cal M}}
\newcommand{\mc}{{\cal M_{\rm cl}}}
\newcommand{\mcn}{{\cal M_{\rm cl}}^{(n)}}
\newcommand{\mn}{{\cal M}^{(n)}}
\newcommand{\tm}{\widetilde{\cal M}}
\newcommand{\tmn}{\widetilde{\cal M}^{(n)}}
\newcommand{\msu}{{\cal M}_{SU}^{(2)}}
\newcommand{\msp}{{\cal M}_{Sp}^{(2)}}

\newcommand{\ff}{{\cal F}}
\newcommand{\tg}{{\widetilde G}}
\newcommand{\ftg}{{\cal F}_{\widetilde G}}

\newcommand{\w}{\omega}

\newcommand{\bZ}{{\bf Z}}
\newcommand{\bC}{{\bf C}}
\newcommand{\bR}{{\bf R}}

\section{Introduction}
S-duality is the quantum equivalence of classically inequivalent field
theories.  The paradigmatic example is the strong-weak coupling
duality of $N=4$ supersymmetric Yang-Mills under which theories
with couplings $\tau$ and $-1/\tau$ are identified.

This letter discusses S-dualities for scale invariant $N=2$
supersymmetric field theories with product gauge groups and matter in
the fundamental representation \cite{w9703,lll9705,bsty9705,
kmv9706,ll9708}.
The main evidence for S-dualities in $N=2$ theories has come from the
spectrum of BPS saturated states \cite{mo77} and from low-energy
effective actions \cite{sw9408}.  By relating S-duality
transformations in scale invariant $N=2$ gauge theories to global
symmetries in asymptotically free theories, it was shown in
\cite{a9706} that S-dualities of gauge theories with a single gauge
group factor are, in fact, {\it exact} equivalences of their quantum
field theories.  S-dualities of the theories with product gauge groups
are expected to have a more complicated structure of identifications
on the classical coupling space \cite{w9703,lll9705,bsty9705,kmv9706,
ll9708}.  Nonetheless, we show
that it is still possible to relate S-dualities of these theories to
global symmetries of higher rank asymptotically free theories.

As for simple gauge groups, the basic idea is to regard the marginal
couplings of the scale invariant theory as the lowest components of
$N=2$ vector multiplets---complex scalar ``Higgs'' fields---in an
enlarged theory.  Then the coupling space $\mm$ of the scale invariant
theory is realized as a submanifold of the Coulomb branch $\cal C$ of
the enlarged theory.  Any S-duality identifications of different
points of $\mm$ are interpreted as equivalences on $\cal C$.  By
choosing the enlarged theory appropriately, these equivalences on
$\cal C$ can be made manifest as (spontaneously broken) global
symmetries.

In the next section we review the derivation of the S-duality
transformations for the gauge theories with simple gauge group
\cite{a9706}.  The brane picture of the derivation is very
illuminating and allows for the straightforward generalization to the
theories with product gauge groups.  In section 3 we consider scale
invariant $\bigotimes_i SU(k_i)$ gauge theories and in section 4 we
derive S-dualities of scale invariant $\bigotimes_i
SO(k_{2i-1}){\otimes} Sp(k_{2i})$, $\bigotimes_i\! SU(k_i) \otimes
SO(n)$ and $\bigotimes_i\! SU(k_i) \otimes Sp(n)$ theories.  We
conclude with some comments in section 5.


\section{Review of S-duality for simple gauge groups}
The scale invariant $N=2$ theories with a single $SU$, $SO$, or $Sp$
gauge group and quarks in the fundamental (defining) representation
all have low-energy effective theories that are invariant under
identifications of $\tau$ under a discrete group isomorphic
to\footnote{For the $SU(2)$ scale invariant theory the S-duality group
is enlarged to $SL(2,\bZ)$ \cite{sw9408}; we return to this point in
section 5.}  $\Gamma^0(2) \subset SL(2,\bZ)$ \cite{aps9505,as9509}.
$\Gamma^0(2)$ is the subset of $SL(2,\bZ)$ matrices with even upper
off-diagonal entry, or equivalently, is the subgroup of $SL(2,\bZ)$
generated by $\widetilde T:\ \tau\to\tau+2$ and $S:\ \tau\to-1/\tau$,
acting on the classical coupling space $\mc = \{\hbox{Im}\tau > 0 \}$.
(We have taken the gauge coupling to be $\tau = {\vartheta \over \pi}
+ i {8 \pi \over g^2}$, differing by a factor of two from the usual
definition.)  $\Gamma^0(2)$ is characterized more abstractly as the
group freely generated by two generators $\widetilde T$ and $S$
subject to one relation $S^2=1$.  Dividing $\mc$ by the S-duality
group gives the the quantum coupling space $\mm = \mc /\Gamma^0(2)$.
Generally, the relations satisfied by the generators $\{\widetilde
T,S\}$ encode the holonomies in $\mm$ around the fixed points of the
$\Gamma^0(2)$ action.  For example, for the action given above, $\mm$
has three such points: the weak coupling point $\tau=+i\infty$ fixed
by $\widetilde T$, an ``ultra-strong'' coupling point $\tau=1$ fixed
by $\widetilde TS$, and a $\bZ_2$ point $\tau =i$ fixed by $S$.  In
fact, this data plus the fact that the topology of $\mm$ is that of a
two sphere summarizes the physically meaningful information about the
space of couplings.  Its particular realization as a fundamental
domain of $\Gamma^0(2)$ acting on the $\tau$ upper half plane is
dependent on which coordinate $\tau$ we use; without an independent
non-perturbative definition of $\tau$, the only conditions we can
physically impose are on its behavior at arbitrarily weak coupling.
Thus the physical content of a statement of S-duality is nothing more
than a characterization of the topology of the space of couplings and
the holonomies around its singular points.

Consider the scale invariant $SU(r)$ theory with $2r$ fundamental 
quarks. The Coulomb branch of the theory is described by the 
curve \cite{aps9505}
\be
y^2 = \prod_{a=1}^{r}(x-\phi_a)^2 + (h^2-1)\prod_{j=1}^{2r}
(x-\mu_j-h\mu),
\label{susingle}
\ee 
parameterized by a gauge coupling $h$, quark masses $\mu_j$ and
$\mu$, and Higgs vevs $\phi_a$.  $h$ is a function of the coupling
such that $h^2 \sim 1 + 64e^{i\pi\tau}$ at weak coupling, $\mu_j$
(satisfying $\sum_j\mu_j=0$) are the eigenvalues of the mass matrix
transforming in the adjoint of the $SU(2r)$ flavor group, the singlet
mass $\mu$ is charged under the $U(1)$ ``baryon number'' global
symmetry, and the $\phi_a$ (satisfying $\sum_a \phi_a =0$) are the
eigenvalues of the adjoint Higgs field; only flavor and gauge
invariant combinations of the $\mu_j$ and $\phi_a$ appear as
coefficients in (\ref{susingle}).

In the scale invariant theory (setting the masses to zero) in this
parameterization of the curve, the coupling parameter space is the $h$
plane.  There are two weak coupling points $h=\pm1$ and a point
$h=\infty$ where the low energy effective theory on the Coulomb branch
is singular.  (Though $h=0$ is apparently also such a singular point,
a more careful analysis of the low energy effective action as $h \to
0$ reveals no divergences.)  The low energy effective action is
invariant under the $\bZ_2$ identification $h \to -h$ (with $\mu \to
-\mu$).  Identifying the $h$-plane under the action of this $\bZ_2$
gives a quantum coupling space $\mm$ with one weak coupling
singularity ($h=1$), a $\bZ_2$ fixed point ($h=0$), and a singular
point ($h=\infty$).  This matches the description given above of the
fundamental domain of $\Gamma^0(2)$ and is the low energy evidence for
this S-duality.

We now review the scaling argument of \cite{a9706} which derives this
S-duality by embedding the scale invariant theory in the
asymptotically free $SU(r+1)$ theory with $2r$ quarks.  We can flow to
the scale invariant $SU(r)$ theory by Higgsing the $SU(r+1)$ theory so
that $\phi_a=M,\ 1\le a \le r$ and $\phi_{r+1} = -rM$ and assigning
the singlet mass $\mu= M$ to keep the $2r$ quarks massless.  These
tunings are also valid in the quantum theory, since upon applying them
to the asymptotically free $SU(r+1)$ curve 
\be 
y^2 = \prod_{a=1}^{r+1} (x-\phi_a)^2 - \Lambda^2 (x-\mu)^{2r}
\label{assympsu}
\ee
and shifting $x \rightarrow x+M$, the curve factorizes as
\be
y^2 =x^{2r}\left[ (x+(r+1)M)^2 - \Lambda^2 \right].
\label{sudsin}
\ee
We recognize the $x^{2r}$ factor of (\ref{sudsin}) as the singularity
corresponding to the scale invariant vacuum of the $SU(r)$ theory with
$2r$ massless quarks.  We identify the dimensionless parameter
$M/\Lambda$ which varies along the scale invariant singularity with
(some holomorphic function of) the gauge coupling of the scale
invariant theory.  Denote by $\cg$ the submanifold (parameterized by
$M$) of scale invariant vacua of the Coulomb branch.  From the
degenerations of (\ref{sudsin}) $\cg$ has a weak coupling point at
$M=\infty$ and two ultra-strong coupling points (singularities) at $M
= \pm \Lambda/(r+1)$.  Furthermore there is a non-anomalous $\bZ_2
\subset U(1)_R$ which acts on the Higgs fields as $\phi \rightarrow
-\phi$ (and when appropriately combined with a global flavor rotation
takes $\mu \rightarrow -\mu$), so that the $M$ plane is identified
under $M \rightarrow -M$,\footnote{Although one should not quotient
the space of vacua by the action of a spontaneously broken global
symmetry, because we are interpreting this submanifold of the Coulomb
branch as the {\it coupling} space of the scale invariant theory, it
is legitimate (indeed necessary) to do so.} giving a single
ultra-strong coupling point and a $\bZ_2$ orbifold point at $M=0$,
thus deriving the content of the conjectured S-duality for the $SU$
series.

Although this argument used the low energy effective action of the
asymptotically free theory, it provides more than just low energy
evidence for the S-duality.  By tuning vevs on the Coulomb branch
$\cc$ of the asymptotically free theory to approach the scale
invariant fixed line $\cg \subset \cc$, we can deduce exact
information about the scale invariant theories.\footnote{This same
type reasoning is used in \cite{m9711} to deduce exact equivalences
between conformal field theories and supergravity theories.}  Since
the manifold of fixed points $\cg$ is related to the coupling space
$\mm$ of the scale invariant theory by a holomorphic map (by $N=2$
supersymmetry), $\cg$ must be a multiple cover of $\mm$, and hence the
topology of $\cg$ will constrain the topology of $\mm$, giving exact
S-duality relations.  These identifications are exact in the the scale
invariant theory because by approaching $\cg$ arbitrarily closely on
the Coulomb branch we can make the effect of any irrelevant operators
from the asymptotically free theory as small as we like.  In essence,
this argument assumes only that the scale invariant theory has a gap
in its spectrum of dimensions of irrelevant operators.\footnote{By the
arguments of \cite{p88} this is plausibly also the condition for the
scale invariant theory to be conformally invariant.}

The explicit factorization of (\ref{sudsin}) into the conformal factor
of the scale invariant $SU(r)$ theory and the factor responsible for
additional singularities (which we interpreted as singularities on the
space of couplings) is possible because of the special hyperelliptic
representation of this curve.  Solutions of $N=2$ gauge theories with
product gauge groups are encoded in curves which are not
hyperelliptic.  To generalize the preceding argument to product gauge
groups it is useful to reformulate it in an M-theory/IIA brane
language.\footnote{In \cite{kmv9706} S-dualities for a broader
class of theories are derived by considering
type II strings on Calabi-Yau three-folds.}

An {\it ad hoc} object from the field theory perspective---a Riemann
surface whose complex structure encodes the low energy couplings on
the Coulomb branch---is given a physical interpretation in the
M-theory formulation of $N=2$ gauge theories \cite{klmvw9604,w9703}.
Consider the type IIA
configuration of intersecting branes depicted in Fig.~1. Two (NS)
fivebranes extend in the directions $x^0,x^1,\dots,x^5$ and are
located at $x^7=x^8=x^9=0$ and at some specific values of $x^6$.
$r+1$ (D) fourbranes stretch between the fivebranes, extend over
$x^0,\dots,x^3$ and are located at $x^7=x^8=x^9=0$ and at a point in
$v=x^4+i x^5$. Note that these fourbranes are finite in the $x^6$
direction. This configuration represents the Coulomb branch vacua of
an $SU(r+1)$ gauge theory in the four dimensions $x^0,\dots,x^3$, with
the position of the fourbranes in $v$ corresponding to the vevs of the
adjoint scalar in the $N=2$ vector multiplet.  For the model in hand
we choose $v=M$ for $r$ of the fourbranes and $v=-r M$ for the
remaining one.  In Fig.~1 we have also shown $2r$ semi-infinite
fourbranes which also extend over $x^0,\dots,x^3$ and are located at
$x^7=x^8=x^9=0$.  $r$ of these fourbranes end on the left fivebrane
and the other $r$ end on the right one and correspond to the $2r$
hypermultiplets (quarks) of the $SU(r+1)$ gauge theory.  Their (fixed)
$v$ coordinates encode the masses of the hypermultiplets.  With the
choice of mass parameters as in (\ref{assympsu}), one gets the
configuration of Fig.~1.

\begin{figure}
\centerline{
\psfig{figure=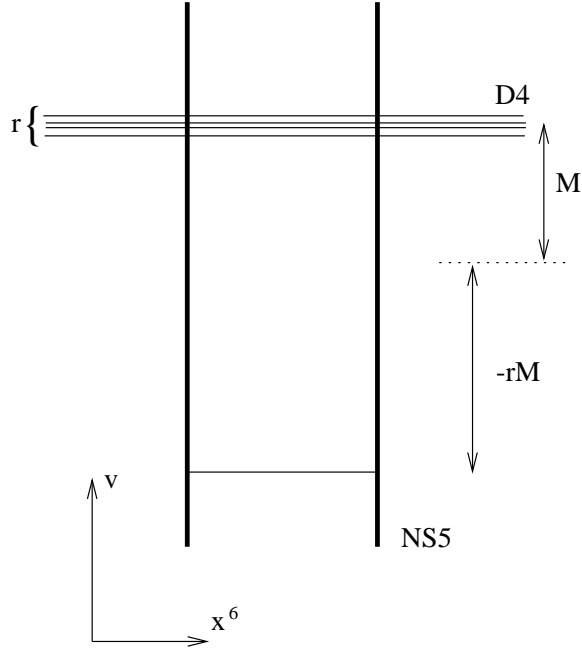,width=3truein}}
{\caption{A configuration of fivebranes connected by parallel 
fourbranes realizing the embedding of the scale invariant $SU(r)$ 
gauge theory into  an asymptotically free $SU(r+1)$ theory. The
horizontal dotted line marks the position of the origin in the 
$v$-plane.}
\label{fr}}
\end{figure}

This brane configuration is singular where the ends of the fourbranes
end on the fivebrane world volume.  These singularities are resolved
by lifting the construction to M theory \cite{w9703} where the IIA
fourbranes become M theory fivebranes wrapped around the eleventh
dimension $x^{10}$. In fact, the whole configuration of branes in
Fig.~1 appears as a single M theory fivebrane with world volume
$\bR^4\times \Sigma$ where $\bR^4$ is the four-dimensional low energy
space-time and $\Sigma$ is the Riemann surface describing the Coulomb
branch of the $SU(r+1)$ gauge theory with $2r$ quarks.  Let $t=e^{-s}$
with $s=(x^6+i x^{10})/R$, $R$ being the radius of the eleventh
dimension.  Then $\Sigma$ is holomorphically embedded in
$\bC^2=\{t,v\}$ as the curve \cite{w9703}
\be
(v-M)^r \ t^2+{\tilde f}\ (v-M)^r (v+r M)\  t+(v-M)^r=0.
\label{mthsu}
\ee
{}From the Type IIA perspective (\ref{mthsu}) describes
bending fivebranes as $t\to \infty$ or $0$.  Asymptotically, 
the separation of the fivebranes reads  
\be
s_2-s_1=\ln {t_1\over t_2}\simeq 2\ln \tilde{f} v 
\label{bend}
\ee
which describes the running of the $\beta$-function of the
asymptotically free $SU(r+1)$ gauge theory with $2r$ quarks, with
$1/{\tilde f}$ identified with the strong coupling scale $\Lambda$ of
the $SU(r+1)$ theory, and $v$ interpreted as the energy scale at which
we measure the coupling.

The scaling limit, implicit in (\ref{sudsin}), involves looking at the
scale invariant $SU(r)$ theory at energy scales much smaller than
either $M$ or $\Lambda$.  Only far below these scales can one sensibly
talk about the scale invariance of the $SU(r)$ gauge theory which
appears as an effective low energy description after integrating out
modes charged under the decoupled $U(1)$ gauge factor.  The decoupling
of the $U(1)$ gauge factor from the brane point of view means that we
consider a small region of the brane construction near $v=M$, which
locally looks like a finite $SU(r)$ gauge theory.  To describe the
scale invariant geometry near $v=M$ we thus introduce a local
coordinate $x=v-M$ and consider $x\ll\{\Lambda,M\}$ which is the
geometrical realization of the scaling to the fixed point theories.
{}From (\ref{mthsu}) we find the description of this region to be
\be 
x^r \left(t^2+{M(r+1)\over\Lambda}\ t +1\right)=0,
\label{mthscale}
\ee
which indeed describes the scale invariant $SU(r)$ theory (at the
origin of its Coulomb branch) with coupling parameter
\be
f=M(r+1)/\Lambda.
\label{f}
\ee

Thus, we have tuned to a one complex dimensional submanifold of the
Coulomb branch $\{M\}\equiv\cg\subset\cc$ of the $SU(r+1)$ theory,
which realizes some multiple cover $\tm$ of the coupling space of the
scale invariant $SU(r)$ theory.  At certain points in $\cg$,
specifically $M=\pm2\Lambda/(r+1)$ and $M\to \infty$, (\ref{mthscale})
develops additional singularities that we interpret through (\ref{f})
as singularities (``punctures'') of $\tm$. $\tm$ is not quite the
coupling manifold $\mm$ of the $SU(r)$ theory since, as in the field
theoretical arguments above, there is a $\bZ_2$ identification $M\to
-M$ coming from the unbroken subgroup of the $U(1)_R$ symmetry:
\be 
{\mm}\ \simeq\ \tm/\bZ_2 .
\label{suquantum}
\ee
This once again gives the expected S-duality of the $SU(r)$ theory.

S-dualities can be derived for the scale invariant $SO$ and $Sp$
theories along the same lines.  One can also provide a similar
interpretation of the derivation in brane language.

Before proceeding to the generalization of this construction to the
product gauge group theories, we comment on the definition of
S-duality {\it groups}.  It is natural to define the S-duality group
$\Gamma$ by $\mm=\mc/\Gamma$.  Using the fact that $\pi_1(\mc)=1$ one
might be tempted to identify $\Gamma = \pi_1(\mm)$, the fundamental
group of $\mm$.  However, as we have just seen, $\Gamma$ does not act
freely on $\mc$, so $\pi_1(\mm)$ does not capture all of $\Gamma$.  In
particular, we must enlarge $\pi_1$ to include the holonomies around
the orbifold singularities of $\mm$ \cite{a9705}.  For example, we can
consider fixed points of discrete identifications as being punctures,
and identify a closed loop around a $\bZ_2$ orbifold singularity $M=0$
of $\mm$ with the order two element of the S-duality group.  Adding to
this the other nontrivial element of the fundamental group of $\mm$
represented by a closed loop around the $M^2=4\Lambda^2/(r+1)^2$
puncture, we recover the full S-duality group $\Gamma^0(2)$ of the
scale invariant $SU(r)$ theory.


\section{S-duality in scale invariant $\bigotimes$SU gauge theories}
In this section we consider the ``cylindrical'' models of \cite{w9703}
and implement their non-perturbative duality group as global symmetries
acting on the Coulomb branch of a higher rank theory.

Consider $N=2$ supersymmetric gauge theory with gauge group 
\be
G=\bigotimes_{i=1}^n SU(k_i),
\label{color}
\ee
$n-1$ hypermultiplets in the bifundamental representation $\bf
(k_1,\overline{k}_2)\oplus(k_2,\overline{k}_3)\dots
\oplus(k_{n-1},\overline{k}_n)$ and $d_i$ hypermultiplets in the
fundamental representation of the $SU(k_i)$ gauge group factor. We
denote this model as 
\be
\ff_G={\bf {k_1}_{\ d_1}|\dots|{k_i}_{\ d_i}|\dots|{k_{n}}_{\ d_n}}.
\ee
Given a gauge group $G$ there is a unique choice of $d_i$ for which 
the beta functions all vanish, namely
\be 
d_i=2k_i-k_{i-1}-k_{i+1}
\label{sclinv}
\ee 
(where we understand $k_0=k_{n+1}=0$).  The Coulomb branch $\cc$ is
the manifold of vacua with zero hypermultiplet vevs and
vector multiplet vevs in the complexified Cartan subalgebra of $G$.
Locally $\cc\simeq\bC^r$, where $r={\rm rank}\ G$, with coordinates
$\phi_j^{(i)},\ j=1,\dots k_i,\ i=1,\dots n$.  The origin of $\cc$ 
($\phi_j^{(i)}=0$ for all $\{j,i\}$) describes the scale invariant 
vacuum, while elsewhere the scale invariance is spontaneously broken
and the underlying theory generically flows to an infrared free 
$U(1)^r$ gauge theory.  The couplings $\tau_{a b}$ of this low-energy
effective $U(1)^r$ theory can be geometrically encoded as the complex
structure (period matrix) of a genus-$r$ Riemann surface $\Sigma$.

Following \cite{w9703} for the given model $\Sigma$ is 
\bea
y^{n+1}&+&g_1(v)y^n+g_2(v)J_1(v)y^{n-1}+g_3(v)J_1(v)^2J_2(v)y^{n-2}\cr
&+&\dots+g_i(v)\prod_{s=1}^{i-1}J_s^{i-s} y^{n+1-i}+\dots+
\prod_{s=1}^{n}J_s^{n+1-s}\ =0
\label{su}
\eea
with 
\bea
J_i&=&\prod_{j=1}^{d_i}(v-m_j^{(i)})\cr
g_i&=&f_i\ \prod_{j=1}^{k_i}(v-\mu_i-\phi_j^{(i)}),\qquad 
\sum_{j=1}^{k_i-1}\phi_j^{(i)}=0 .
\label{su1}
\eea
Here $m_j^{(i)}$ are masses of the hypermultiplets in the fundamental
representation of the $j$th gauge group factor, $\mu_{i+1}-\mu_i$ are
the masses of the hypermultiplets in the bifundamental representations
$\bf(k_i,\overline{k}_{i+1})$, and the $f_i$ are some functions of
the $n$ gauge couplings $\tau_i$ of the different factors in $G$.
The dependence of the coefficients of the polynomial (\ref{su}) on
the bares masses, vevs, and couplings can be determined by taking
various weak coupling limits of the above curve (corresponding in the
IIA picture to taking groups of NS fivebranes
off to infinity), and matching to known lower rank results.  This also
determines the $f_i$ as functions of the $\tau_i$ as (though we will
not need this result in what follows)
\be
f_i = \bigl( \sum_{j_1>\cdots>j_i} s_{j_1}\cdots
s_{j_i} \bigr) \left(\prod_{i=1}^{n+1} s_i\right)^{-i/(n+1)} ,
\ee
where $s_i\equiv-\prod_{k=1}^{i-1} q_k$ (and $s_1\equiv-1$) and $q_i =
e^{i\pi\tau_i}$.  Note that all these identifications (of the masses
and vevs as well as the couplings) can be modified by adding terms
which vanish at least as fast as a power law in the $q_i$ at weak
coupling ($q_i\to0$).  Such redefinitions just correspond to
non-perturbative redefinitions of the parameters which we have no
independent way of fixing.  The dependence of the coefficients of
polynomials describing the curves for non-scale invariant
(asymptotically free) theories can be found from the above
identifications by appropriately sending bare masses to infinity and
couplings to zero.  We used such parameter matchings to write equation
(\ref{mthsu}) for the simple $SU$ theory considered in the last
section, and we will use such matchings several times more in the
remainder of the paper.  Some examples of these identifications had
been derived previously in \cite{enr9801}.

The classical space of couplings $\mcn$ is $\bC^n$ corresponding to
the $n$ gauge couplings $q_i$.  Just as in the case of a single gauge
group the ``quantum'' coupling space $\mn$ seen by the low energy
effective theory---as encoded in the curve (\ref{su})---is quite
different.  It can be identified with the asymptotic positions of the
M theory fivebrane on the cylinder parameterized by the complex
coordinate $s=(x^6+ix^{10})/R$.  This identification can be expressed
following \cite{w9703} as
\be
\mn = {\mm}_{0,n+3;2}
\label{toshow}
\ee
where ${\mm}_{0,n+3;2}$ is the moduli space of a genus zero Riemann
surface with $n+3$ marked points, two of which are distinguished and
ordered while the other $n+1$ are indistinguishable.  (In \cite{w9703}
the S-duality group $\Gamma_n$ for these theories was identified with
$\pi_1({\mm}_{0,n+3;2})$.  This must be extended to include holonomies
around the orbifold points as discussed in the last section.)  In what
follows we will present a scaling argument to derive this S-duality
as an exact equivalence by deriving (\ref{toshow}).

We start with an asymptotically free theory described in the
brane language by
\be
\ftg={\bf {k_1+1}_{\ d_1}|\dots|{k_i+i}_{\ d_i}|\dots|
{k_{n}+n}_{\ d_n}}.
\ee
We can flow to a scale invariant $\ff_G$ model by tuning the Higgs
vevs $\widetilde{\phi}_j^{(i)}$ of the $\ftg$ model so as to break
\be
\tg=\bigotimes_{i=1}^{n} SU(k_i+i)\to \bigotimes_{i=1}^{n}U(1)^i
\ \otimes\ G.
\label{gb}
\ee
Classically, this is achieved by choosing the Higgs vevs of
$\ff_{\tg}$ as
\bea
\phi_{j}^{(1)}=M_1,\ j=1,\dots,k_1\qquad &{\rm and}&\qquad 
\phi_{k_1+1}=-k_1 M_1\cr
\phi_{j}^{(2)}=M_1,\ j=1,\dots,k_2\qquad &{\rm and}&\qquad 
\phi_{k_2+p}=-k_2 M_1+\w_2^p M_2,\ p=1,2\cr
\dots\qquad\qquad\qquad &{}&\qquad\qquad\qquad\dots\cr
\phi_{j}^{(i)}=M_1,\ j=1,\dots,k_i\qquad &{\rm and}&\qquad 
\phi_{k_i+p}=-k_i M_1+\w_i^p M_i,\ p=1,\dots,i\cr\label{suvev}
\dots\qquad\qquad\qquad &{}&\qquad\qquad\qquad\dots\cr
\phi_{j}^{(n)}=M_1,\ j=1,\dots,k_n\qquad &{\rm and}&\qquad 
\phi_{k_n+p}=-k_n M_1{+}\w_n^p M_n,\ p=1,\dots,n
\eea
where $\w_p$ is the $p$th root of unity, $\w_p=e^{2\pi i/p}$.
In addition, to insure that all hypermultiplets of the $\ff_G$ 
model are massless we choose zero bifundamental masses of $\ftg$, 
$\widetilde\mu_i=0$, $i=1,\dots,n$, and tune all masses of the 
hypermultiplets in the fundamental representations of the $\tg$ 
factors to ${\widetilde m}_{j}^{(i)}=M_1$, $j=1,\dots,d_i$;
$i=1,\dots,n$.

These tunings are also valid quantumly at energy scales much lower
than ${\rm min}_i M_i$. To see this, consider the curve
$\widetilde\Sigma$ for the $\ftg$ model
\bea
y^{n+1}&+&{\tilde f}_1 {(v-M_1)}^{k_1}(v+k_1 M_1)y^n\cr 
&+&{\tilde f}_2 {(v-M_1)}^{k_2}(v-M_1)^{d_1}y^{n-1}\prod_{p=1}^2(v+
k_2 M_1-\w_2^p M_2 )+\dots\cr
&+&{\tilde f}_i {(v-M_1)}^{k_i}(v-M_1)^{\sum_{s=1}^{i-1}
d_s(i-s)}y^{n+1-i}
\prod_{p=1}^i(v+k_i M_1-\w_i^p M_i )+\dots\cr
&+&{(v-M_1)}^{\sum_{s=1}^{n}d_s(n+1-s)}=0\ .
\label{sua}
\eea
Shifting $v\to v+M_1$, in the decoupling limit,
$v\ll {\rm min}_i M_i$,  $\widetilde\Sigma$ turns into 
\bea
y^{n+1}&+&{\tilde f}_1 (k_1+1)M_1 v^{k_1}y^n+\dots\cr
&+&{\tilde f}_i \big((k_i+1)^i M_1^i-M_i^i\big) v^{k_i}
\prod_{s=1}^{i-1}{\big(v^{d_s}\big)}^{i-s}y^{n+1-i}+\dots\cr
&+&\prod_{s=1}^{n}{\big(v^{d_s}\big)}^{n+1-s}=0,
\label{sud}
\eea   
which we recognize as the curve for the scale invariant $\ff_G$ 
model (\ref{su}) with gauge coupling parameters
\be
f_i=\tilde{f}_i \big((k_i+1)^i M_1^i-M_i^i\big).
\label{couplings}
\ee
Note that the first $n-1$ gauge factors of the $\ff_{\tg}$ model have
zero beta function, while the last one is asymptotically free.  We
therefore must assign a strong coupling scale $\Lambda$ to this
factor. By embedding the scale invariant model $\ff_G$ into
$\ff_{\tg}$ we effectively traded the $n$ dimensionless couplings
$q_i$ of $\ff_G$ for the $n$ gauge invariant dimensionless parameters
$x_i=(M_i/\Lambda)^i$ of the $\ff_{\tg}$ model. Let
\be
\tmn=\{ x_i,\ i=1,\dots,n\}.
\label{coversu}
\ee   
At special points in $\tmn$, (\ref{sud}) develops addition
singularities that should be thought of as the singularities in the
quantum coupling space.  This locus of singularities (punctures) in
$\tmn$ is easily extracted from (\ref{sud}) by redefining $y\to
v^{k_1} y$ and using the scale invariance conditions (\ref{sclinv})
to get
\be
y^{n+1}+{\tilde f}_1 (k_1+1)x_1 y^n+\dots
+{\tilde f}_i \big((k_i+1)^i x_1^i-x_i\big) y^{n+1-i}+\dots
+1=0,
\label{susingularities}
\ee
and is given by the vanishing of the discriminant of
(\ref{susingularities}).  Singularities of (\ref{sud}) also arise when
one of the roots of (\ref{susingularities}) goes either to zero or
infinity, which signals the decoupling of the gauge group factors.

Global symmetries of the $\ftg$ model induce identifications on the
$\tmn$ manifold.  The classical $U(1)_R$ symmetry of $\ftg$ is broken
by the instantons of the last factor in $\tg$ to a discrete subgroup
$\bZ_{n+1}= \{\w_{n+1}^p,\ p=0,\dots,n\}$.  The latter acts on the
coordinates $x_i$ as
\be
x_i\to \w_{n+1}^{p\cdot i}\  x_i,\qquad p=0,\dots,n,
\label{suaction}
\ee
under whose action we identify the quantum coupling space $\mn$ with 
\be
\mn\ \simeq\ \tmn/{\bZ_{n+1}} .
\label{suid}
\ee
The punctures of the $\tmn$ manifold comprise the vanishing locus of
the discriminant of (\ref{susingularities}) along with points where
one or more roots of (\ref{susingularities}) go to zero or infinity.

We can now show that $\mn\simeq\mm_{0,n+3,2}$.  It is convenient
to change coordinates $\{x_i\}$ on $\tmn$ to $\{f_i\}$ through
(\ref{couplings}).  This change is nonsingular and one-to-one. The
$\bZ_{n+1}$ symmetry of $\tmn$ has the same action on $f_i$ as on
$x_i$.  The polynomial determining the punctures then reads
\be
y^{n+1}+f_1 y^n+\dots+f_i y^{n+1-i}+\dots+f_n y+1=0 .
\label{pf}
\ee
Now, a configuration of $n+1$ unordered points on a genus zero Riemann
surface with two marked points (zero and infinity)---a point in
$\mm_{0,n+3,2}$---can be thought of as the positions of the roots of
the polynomial (\ref{pf}) which are encoded in the coefficients
$\{f_i\}$.  Furthermore, the $n+1$ sets of coefficients
$\{\w_{n+1}^p\cdot f_i\}$, $p=0,\dots,n$ are related by a rotation of
the sphere (which preserves the special character of zero and
infinity) and so should be identified.  Dilatations of $y$ are already
fixed by the constant term in (\ref{pf}).  The set $\{f_i\}$ maps to a
single point in $\tmn$, while the above identification is equivalent
to the modding by $\bZ_{n+1}$ in $\tmn$. Thus,
\be
\mn = \tmn/{\bZ}_{n+1}\ \simeq\ {\mm}_{0,n+3,2} ,
\ee
as we sought to prove.
 
We can also show how decoupling of a single gauge
factor in $G$ produces ${\mm}^{(n-1)}$ from $\mn$.  Decoupling the
first gauge factor in the product chain is achieved by taking all
$f_i\to \infty$, while keeping
\be
{\tilde{p}}_i={f_{i+1}\over{f_1^{1-i/n}}}
\label{tif}
\ee 
finite.  In this limit, $\tmn\to {\widetilde{\cal P}}^{(n-1)}$ with
the punctures of the ${\widetilde{\cal P}}^{(n-1)}$ manifold described 
by
\be
y^{n}+p_1 y^{n-1}+\dots+p_i y^{n-i}+\dots+p_n y+1=0 ,
\label{pp}
\ee
which follows from (\ref{pf}) in the limit after rescaling $y\to y/
f_1^{1/n}$.  The $\bZ_{n+1}$ action on $\tmn$ induces a $\bZ_n$ action
on ${\widetilde{\cal P}}^{(n-1)}$
\be
p_i\to \w_{n+1}^{p\cdot i\cdot (1+1/n)}\ p_i\ =\ \w_n^{p\cdot i} p_i.
\label{actionp}
\ee
Thus we conclude that indeed ${\widetilde{\cal P}}^{(n-1)}\simeq
{\tm}^{(n-1)}$.


\section{S-duality in other product gauge groups}
$N=2$ gauge theories with products of $SO$ and $Sp$ groups were solved
in \cite{lll9705,bsty9705}. The derivation of S-dualities for these
models essentially repeats the arguments of the previous section.

The model
\be
\ff_G={\bf +{2k_1}_{\ d_1}|-{2k_2}_{\ d_2}|\dots|(-1)^i 
{2k_i}_{\ d_i} |\dots|(-1)^n {2k_{n}}_{\ d_n}}
\label{suso}
\ee
(where ``$+$'' signs denote $SO$ gauge factors and ``$-$'' signs
represent $Sp$ gauge factors) is scale invariant if
\be
d_i=2k_i+ 2(-1)^i -k_{i-1}-k_{i+1}, \qquad i=1,\dots,n,
\qquad k_0\equiv k_{n+1} \equiv 0.
\ee
It can be obtained by Higgsing the
\bea
\ff_{\tg}={\bf + {2(k_1+1)}_{\ d_1}| - {2(k_2+2)}_{\ d_2}|
\dots|(-1)^i {2(k_i+i)}_{\ d_i}
|\dots|(-1)^n {2(k_{n}+n)}_{\ d_n}}
\label{susoa}
\eea
model with all hypermultiplets massless so that 
\be
\tg\to\bigotimes_{i=1}^n U(1)^i\ \otimes\ G .
\label{groupsosp}
\ee
This is achieved by tuning
\bea
&&\widetilde{\phi}_j^{(i)}=0,\qquad
j=1,\dots,k_i;\qquad i=1,\dots,n\cr
&&\widetilde{\phi}_{k_i+1}^{(i)}=M_1;\qquad i=1,\dots,n\cr
&&\dots\cr
&&\widetilde{\phi}_{k_i+r}^{(i)}=M_r;\qquad i=r,\dots,n\cr
&&\dots\cr
&&\widetilde{\phi}_{k_n+n}^{(i)}=M_n.
\label{sospvevs}
\eea
The curve $\widetilde{\Sigma}$ for the $\ff_{\tg}$ model then 
reads \cite{lll9705}
\bea
v^2 y^{n+1}&+&\tilde{f}_1 v^{2k_1} (v^2-M_1^2)\ y^n\cr
&+&\tilde{f}_2 v^{2k_2+2}(v^2-M_1^2)(v^2-M_2^2)v^{2d_1}\ y^{n-1}\cr
&+&\dots\cr
&+&\tilde{f_i} v^{2k_i+1+(-1)^{i}}\ \prod_{r=1}^{i}(v^2-M_r^2)\
v^{2\sum_{s=1}^{i-1}d_s(i-s)}\ y^{n+1-i}\cr
&+&\dots\cr
&+&v^{1+(-1)^{n+1}} v^{{2\sum_{s=1}^{n}d_s(n+1-s)}}\ =0 .
\label{curvesosp}
\eea
In the decoupling limit $v^2\ll M_r^2$, $\widetilde\Sigma$ reproduces
the curve of the scale invariant $\ff_G$ model with gauge couplings
\be
f_i=(-1)^i\tilde{f}_i \prod_{r=1}^i M_r^2 .
\label{couplingsosp}
\ee
As for the product of $SU$ gauge groups, $n-1$ factors of the
$\ff_{\tg}$ model have zero beta function and the last one is
asymptotically free. The instantons of the asymptotically free factor
generate the strong coupling scale $\Lambda$ which breaks the global
$U(1)_R$ symmetry to a $\bZ_{n+1}$ subgroup.  The latter acts on the
dimensionless gauge invariant coordinates $M_i^2$ as $M_i^2\to
\w_{n+1}^p M_i^2$, and hence identifies the couplings under $f_i\to
\w_{n+1}^p\ f_i$, for $p=0,\dots n$.  We thus identify the the quantum
coupling space $\mn$ of the scale invariant $\ff_G$ model with
$\mn\simeq\tmn/{\bZ_{n+1}}$ where $\tmn=\{f_i\}$.  Punctures on $\tmn$
are identified with the points of vanishing discriminant of the
polynomial (extracted from the $\ff_G$ curve by the change of
variables $y\to v^{2k_1-2}y$)
\be
y^{n+1}+f_1 y^n+\dots+f_i y^{n+1-i}+\dots+1=0 
\label{dissossp}
\ee
along with values of $f_i$ which where one or more roots of
(\ref{dissossp}) go to zero or infinity.  This is precisely the same
description of $\mn$ as we found in the last section, and so gives
$\mn\simeq\mm_{0,n+3,2}$, which again is the answer expected on the
basis of the low energy effective theory.

In fact, we find that all the scale invariant models studied in
\cite{lll9705} have the coupling space ${\mm}_{0,n+3,2}$, determined
only by the number of gauge group factors.  This is not surprising
since the $N=2$ gauge theory with $n$ factors in the gauge group was
derived in \cite{lll9705} from a configuration with $n+1$ straight NS
fivebranes which can be associated \cite{w9703} with $n+1$ unordered
points on a cylinder.  The moduli space of these points have
``punctures'' (where the low energy effective description of the
theory is singular) whenever two or more NS fivebranes coincide or
move to infinity.  This can be cast in the form of the vanishing of
the discriminant of an $n+1$ order polynomial as in (\ref{dissossp}).

Unlike the previous example, the $\bigotimes_i\! SU(k_i)\otimes SO(N)$
and $\bigotimes_i\! SU(k_i)\otimes Sp(N)$ series of \cite{ll9708} have
a different quantum coupling space (and therefore a different
S-duality group).  We will now show that in these cases the coupling
space is $\mn={\cal Q}_{0,n,2}$, where ${\cal Q}_{0,n,2}$ is the
moduli space of genus zero Riemann surfaces with two distinguished and
marked points and $n$ unordered pairs of points with the further
condition that, in terms of any complex coordinate $z$ on the sphere,
the product of the coordinates within each pair is the same for all
pairs.
  
We will write out the derivation only for the $\bigotimes_i\! SU(k_i)
\otimes SO(k_n)$ model with even $k_n$, since the other cases can be
analyzed in a similar way.  We flow to the scale invariant
model
\be
\ff_G={\bf {k_1}_{\ d_1}|\dots|{k_i}_{\ d_i}
|\dots|{{k_{n-1}}_{\ d_n-1}|{+k_n}_{\ d_n}}}
\label{sso}
\ee  
with
\be
d_i=2k_i-k_{i-1}-k_{i+1}, \qquad i=1,\dots,n-1,
\qquad k_0\equiv 0, \ {\rm and}\ d_n=k_n-2-k_{n-1},
\ee
by Higgsing the 
\be
\ff_{\tg}={\bf {k_1+2}_{\ d_1}|\dots|{k_i+2i}_{\ d_i}
|\dots|{{k_{n-1}+2n-2}_{\ d_n-1}|{+(k_n+2n)}_{\ d_n}}}
\label{ssoa}
\ee
model with massless hypermultiplets, breaking the gauge
group $\tg$ as
\be
\tg\to \bigotimes_{i=1}^{n-1} U(1)^{2i}\ \otimes U(1)^{n}
\ \otimes\ G.
\label{ssogroup}
\ee 
This is achieved by tuning
\bea
&&{\widetilde\phi}_{j}^{(i)}=0,\qquad j=1,\dots,k_i;
\qquad i=1,\dots,n-1\cr
&&{\widetilde\phi}_{j}^{(n)}=0,\qquad j=1,\dots,k_n/2\cr
&&{\widetilde\phi}_{k_i+1}^{(i)}
=-{\widetilde\phi}_{k_i+2}^{(i)}=M_1;
\qquad i=1,\dots,n-1\cr
&&{\widetilde\phi}_{k_n/2+1}^{(n)}=M_1;\cr
&&\dots\cr
&&{\widetilde\phi}_{k_i+2r-1}^{(i)}
=-{\widetilde\phi}_{k_i+2r}^{(i)}=M_r;
\qquad i=r,\dots,n-1\cr
&&{\widetilde\phi}_{k_n/2+r}^{(n)}=M_r;\cr
&&\dots\cr
&&{\widetilde\phi}_{k_{n-1}+2n-1}^{(n-1)}=
-{\widetilde\phi}_{k_{n-1}+2n}^{(n-1)}=M_n;\cr
&&{\widetilde\phi}_{k_n/2+n}^{(n)}=M_n\ .
\label{ssovevs}
\eea
Introducing 
\bea
&&g_i(v)=g_{2n-i}(-v)=\tilde{f}_i v^{k_i}
\prod_{r=1}^{i} (v^2-M_r^2),\ i=1,\dots,n\cr
&&J_i(v)=J_{2n-i}(-v)=v^{d_i},\ i=1,\dots,n-1\cr
&&J_n(v)=(-1)^{d_n}v^4\ v^{2d_n} ,
\label{shothand}
\eea
the curve $\widetilde{\Sigma}$ for the $\ff_{\tg}$ model 
becomes \cite{ll9708}
\bea
y^{2n}+g_1(v) y^{2n-1}+\dots+
g_i(v)\prod_{r=1}^{i-1}J_r^{i-r} y^{2n-i}+\dots+
\prod_{r=1}^{2n-1}J_r^{2n-r}
=0 .
\label{ssocurve}
\eea
In the decoupling limit $v^2\ll M_r^2$ it reduces 
to the scale invariant 
$\ff_G$ model with gauge couplings
\be
f_i=f_{2n-i}=(-1)^i \tilde{f}_i \prod_{r=1}^{i}M_r^2,\qquad 
i=1,\dots,n.
\label{gaugesso}
\ee
The $SU$ factors of the $\tg$ group have vanishing beta function and
the $SO$ factor is asymptotically free. The global $U(1)_R$ symmetry
is broken to a $\bZ_2$ discrete subgroup which acts on the gauge
invariant parameters $M_i^2$ as $M_i^2\to -M_i^2$, and hence
identifies the couplings by $f_i\to -f_i$, $i=1,\dots,n$.  The quantum
coupling space $\mn$ of the scale invariant $\ff_G$ model is thus
$\mn\simeq\tmn/\bZ_2$ where $\tmn=\{f_i\}$.  Punctures on $\tmn$ are
at values of $f_i$ for which either the discriminant of the polynomial
(extracted from the curve of the $\ff_G$ model by the change of
variables $y\to v^{k_1}y$)
\be
y^{2n}+f_1 y^{2n-1}+\dots+f_{n-1}y^{n+1}+f_n y^n
+f_{n-1}y^{n-1}+\dots+f_1 y+1=0
\label{puncturesso}
\ee
vanishes or some or its roots go to zero or infinity.  It is always
possible to factorize a polynomial of the form (\ref{puncturesso})
into one of the form
\be
\prod_{i=1}^n (y^2+z_i y+1)=0,
\label{alpol}
\ee
thus giving an isomorphism between $\tmn=\{f_i\}$ and
$\widetilde{\cal Q}^{(n)}=\{z_i$ modulo permutations$\}$.
The $\bZ_2$ identification $f_i\to-f_i$ becomes $z_i\to -z_i$.  With
this presentation it is clear that $\mn = \widetilde{\cal
Q}^{(n)}/\bZ_2 \simeq {\cal Q}_{0,n,2}$.


\section{Conclusion}
In this letter we have derived the S-dualities of scale invariant
$N=2$ gauge theories with product gauge groups by embedding those
theories in asymptotically free $N=2$ theories with higher rank gauge
groups and tuning to the scale invariant theories on the Coulomb
branch.  We related S-duality transformations on the couplings of the
scale invariant theories to global symmetries acting on the Coulomb
branch of the higher rank theories.  Since these global symmetries are
exact in the asymptotically free theories, this shows these
S-dualities as exact equivalences of the scale invariant theories and
not just as a property of their supersymmetric states.

The uniform way in which the complicated quantum coupling spaces of
the $N=2$ scale invariant theories were derived by this scaling
procedure suggests that it should be effective more generally.  In
particular it would be interesting to extend this argument to study
S-dualities in scale invariant $N=1$ theories, some classes of which
have been proposed and studied in \cite{ls9503,ks9802}.

Another open question involves the physics of the ``ultra-strong''
coupling points in the coupling spaces of $N=2$ theories.  Unlike the
$N=4$ theories where the $SL(2,\bZ)$ S-duality identifies all the
ultra-strong (Im$\tau = 0$) points with the weak coupling limit of the
theory, we have seen above that the S-dualities of the $N=2$ scale
invariant theories are generically smaller, and leave ultra-strong
points (or manifolds) in their coupling spaces.  One counterexample is
the scale invariant $SU(2)$ theory with four fundamental quark
hypermultiplets, studied in \cite{sw9408}.  There it was found that
the theory in fact has the whole $SL(2,\bZ)$ duality and no
ultra-strong points.  This could also be derived through scaling
arguments \cite{a9706} by comparing the embeddings of $SU(2)$ into
$SU(3)$ and $Sp(4)$ asymptotically free theories.  One should note
that the scaling arguments presented in this paper do not claim to
capture all possible S-dualities---there may be further
identifications of the coupling space which are missed since our
arguments only show that $\mn$ is some multiple cover of the true
coupling space.

This leaves open the possibility that there are further
identifications of the coupling spaces of the $N=2$ theories, perhaps
relating the ultra-strong coupling points to some other weakly coupled
physics.  One place where we know such further identifications must
exist are in theories with $SU(2)$ gauge group factors: for in the
limit that the other factors decouple, the $SU(2)$ factors must have
the full $SL(2,\bZ)$ duality of \cite{sw9408}.  The new S-dualities in
these theories will be explored in \cite{ab9804}.

\section*{Acknowledgments}
It is a pleasure to thank A. Shapere, G. Shiu, H. Tye,
Y. Vtorov-Karevsky, and P. Yi for helpful discussions.  This work is
supported in part by NSF grant PHY-9513717.  The work of PCA is
supported in part by an A.P. Sloan fellowship.


\begin{thebibliography}{99}

\bibitem{w9703} E. Witten, 
{\sl Solutions of four-dimensional field theories via M theory},
hep-th/9703166, \NP{\bf B500}(1997)3.

\bibitem{lll9705} K. Landsteiner, E. Lopez, and D.A. Lowe,
{\sl N=2 supersymmetric gauge theories, branes and orientifolds},
hep-th/9705199, \NP{\bf B507}(1997)197.

\bibitem{bsty9705} A. Brandhuber, J. Sonnenschein, S. Theisen, and
S. Yankielowicz, {\sl M theory and Seiberg-Witten curves: orthogonal
and symplectic groups}, hep-th/9705232, \NP{\bf B504}(1997)175.

\bibitem{kmv9706} S. Katz, P.Mayr, and C. Vafa, {\sl Mirror symmetry
and exact solution of 4d N=2 gauge theories--I}, hep-th/9706110,
\ATMP{\bf 1}(1998)53.

\bibitem{ll9708} K. Landsteiner and E. Lopez, {\sl New curves from
branes}, hep-th/9708118.

\bibitem{mo77} C. Montonen and D. Olive, {\sl Magnetic monopoles as
gauge particles?}, \PL{\bf B72}(1977)117; E. Witten and D. Olive, {\sl
Supersymmetry algebras that include topological charges}, \PL{\bf
B78}(1978)97; H. Osborn, {\sl Topological charges for N=4
supersymmetric gauge theories and monopoles of spin 1}, \PL{\bf
B83}(1979)321; A. Sen, {\sl Dyon-monopole bound states, self-dual
harmonic forms on the multi-monopole moduli space, and SL(2,Z)
invariance in string theory}, hep-th/9402032, \PL{\bf B329}(1994)217.

\bibitem{sw9408} N. Seiberg and E. Witten, {\sl Monopoles, duality and
chiral symmetry breaking in N=2 supersymmetric QCD}, hep-th/9408099,
\NP{\bf B431}(1994)484.

\bibitem{a9706} P.C. Argyres, {\sl S-duality and global symmetries in
N=2 supersymmetric field theory}, hep-th/9706095, to appear in \ATMP

\bibitem{aps9505} P.C. Argyres, M.R. Plesser, and A.D. Shapere, {\sl
Coulomb phase of N=2 supersymmetric QCD}, hep-th/9505100, \PRL{\bf
75}(1995)1699.

\bibitem{as9509} P.C. Argyres and A.D. Shapere, {\sl The vacuum
structure of N=2 super-QCD with classical gauge groups},
hep-th/9509175, \NP{\bf B461}(1996)437.

\bibitem{m9711} J. Maldacena, {\sl The large N limit of superconformal
field theories and supergravity}, hep-th/9711200.

\bibitem{p88} J. Polchinski, {\sl Scale and conformal invariance in
quantum field theory}, \NP{\bf B303}(1988)226.

\bibitem{klmvw9604} A. Klemm, W. Lerche, P. Mayr, C. Vafa, and
N. Warner, {\sl Selfdual strings and N=2 supersymmetric field theory},
hep-th/9604034, \NP{\bf B477}(1996)746.

\bibitem{a9705} P.C. Argyres, {\sl Duality in supersymmetric field
theories}, hep-th/9705076, \NP B (Proc.\ Suppl.) {\bf 61A}(1998)149.

\bibitem{enr9801} J. Erlich, A. Naqvi, and L. Randall, {\sl The
coulomb branch of N=2 supersymmetric product group theories from
branes}, hep-th/9801108.

\bibitem{ls9503} R. Leigh and M. Strassler, {\sl Exactly marginal
operators and duality in four-dimensional N=1 supersymmetric gauge
theory}, hep-th/9503121, \NP {\bf B447}(1995)95.

\bibitem{ks9802} S. Kachru and E. Silverstein, {\sl 4d conformal field
theories and strings on orbifolds}, hep-th/9802183; A. Lawrence, N.
Nekrasov, and C. Vafa, {\sl On conformal field theories in four
dimensions}, hep-th/9803015; M. Bershadsky, Z. Kakushadze, and
C. Vafa, {\sl String expansion as large N expansion of gauge
theories}, hep-th/9803076; Z. Kakushadze, {\sl Gauge theories from
orientifolds and large N limit}, hep-th/9803214.

\bibitem{ab9804} P.C. Argyres and A. Buchel, {\sl New S-dualities in
N=2 theories with SU(2)xSU(2) gauge groups}, to appear.

\end{thebibliography}
\end{document}